\documentclass[fleqn,usenatbib]{mnras}
\usepackage[T1]{fontenc}
\usepackage{ae,aecompl}
\usepackage[latin9]{inputenc}
\usepackage{txfonts}
\usepackage{mathrsfs}
\usepackage{amssymb}
\usepackage{graphicx}
\usepackage{wasysym}

\title[Close encounters in NS binaries]{Neutron star natal kicks:
Collisions, $\mu$TDEs, faint SNe, GRBs and GW sources with preceding
electromagnetic counterparts}

\author[Michaely et al.]{
Erez Michaely,\thanks{E-mail: erezmichaely@gmail.com}
Dimitry Ginzburg
and Hagai B. Perets
\\
Physics Department, Technion - Israel Institute of Technology, Haifa
3200004, Israel\\
}

\date{Accepted XXX. Received YYY; in original form ZZZ}

\pubyear{2016}

\begin{document}
\label{firstpage}
\pagerange{\pageref{firstpage}--\pageref{lastpage}}
\maketitle
\begin{abstract}

Based on the observed high velocity of pulsars it is thought that
neutron stars (NSs) receive a significant velocity kick at birth.
Such natal kicks are considered to play an important role in the the
evolution of binary-NS systems. The kick given to the NS (together
with the effect of mass loss due to the supernova explosion of the
NS progenitor) may result in the binary disruption or lead to a significant
change of the binary orbital properties. Here we explore in detail
the dynamical aftermath of natal kicks in binary systems, determine
their possible outcomes and characterize their relative frequency,
making use of analytic arguments and detailed population synthesis
models. In a fraction of the cases the kick may cast the NS in such
a trajectory as to collide with the binary companion, or pass sufficiently
close to it as to disrupt it (micro tidal disruption event; $\mu$TDE),
or alternatively it could be tidally-captured into a close orbit,
eventually forming an X-ray binary. We calculate the rates of direct
post-kick physical collisions and the possible potential production
of Thorne-Zytkow objects or long-GRBs through this process, estimate
the rates X-ray binaries formation and determine the rates of $\mu$TDEs
and faint supernovae from white dwarf disruptions by NSs. Finally we suggest that natal kicks can produce BH-NS binaries with very short gravitational-wave merger time, possibly giving rise to a new type of promptly appearing eLISA gravitational wave (GW) sources, as well
as producing aLIGO binary-merger GW sources with a unique (likely type Ib/c) supernova
electromagnetic counterpart which precedes the GW merger. 
\end{abstract}


\section{Introduction}

Neutron stars (NSs) are born in supernova (SN) explosions of massive
stars with initial mass of $\gtrsim8M_{\odot}$ \citep{Smartt2009,Smartt2015}.
Some of these are pulsars which can be observed through their frequent
radio pulses. Proper motion measurements of pulsar velocities have
indicated that newly born NSs receive a large natal kick reaching
up to $\sim1,500{\rm kms^{-1}}$ \citep{Hobbs2005} . The origin of
the natal kick is still debated, but it's typically thought to be the
results of an asymmetric explosion of the SN. 

Several studies focused on identifying the natal kick velocity distribution,
by measuring the proper motion of isolated pulsars. \citet{Arzoumanian2002}
modeled the kick velocity distribution as two overlapping Gaussians,
the first with low characteristic velocity of $\sigma_{l}\approx90{\rm kms^{-1}}$,
consisting of $40\%$ of the pulsars, and the second with high characteristic
velocity of $\sigma_{h}\approx500{\rm kms^{-1}}$. Later, \citet{Hobbs2005},
using a sample of 73 young pulsars, modeled the velocity distribution
as a Maxwellian with a velocity dispersion of $\sigma\approx270{\rm kms^{-1}}$,
that corresponds to a mean velocity of $\approx430{\rm kms^{-1}}$.
Recent work done by \citet{Beniamini2016} on double NSs suggests
that the observations are consistent with two explosion mechanisms.
The first, associated with large mass loss up to $\sim2.2M_{\odot}$
and high natal kick velocity. The second, associated with small mass
loss $\apprle0.5M_{\odot}$ with low natal kick velocities. 

Over $80\%$ of the massive O and B stars reside in binaries or higher
multiplicity systems, and when accounting for observational biases
the multiplicity fraction is consistent with $100\%$ \citep{Sana2012,Sana2014,Moe2016}.
\citet{Moe2016} showed that up to a semi major axis (sma) of $\sim50{\rm AU}$
all massive stars are in binaries. Therefore a non-negligible fraction
of all NSs will host a companion in the binary systems, and go through
a post SN interaction between the newly born NS and its companion.
This can result in several possible outcomes: a direct physical collision,
a disruption of the companion due to a very close passage near the
NS (a micro tidal disruption event; $\mu$TDE; \citet{Perets2016}),
a tidal capture into a close orbit, or even significant gravitational
wave emission through a close approach of the NS to another compact
object. Previous studies have discussed some aspects of this problem
(e.g. \citealt{Leonard1994,Troja2010,Hills1983}),
as we discuss below; here we focus on characterizing and mapping all
of these possible outcomes. 

In this work we calculate the possible interaction between the newly
born NS and its binary companion. After the formation of the NS it
may pass sufficiently close to its stellar or Black Hole (BH) companion
to strongly interact with it and produce an electromagnetic
transient, a gravitational wave source or evolve to become a short-period
binary to eventually form an X-ray binary system. The exact dynamical
outcome of the SN explosion is determined by the natal kick velocity
vector, the mass loss in the SN explosion, the initial Keplerian orbit
and the companion radius, which are determined by the long term stellar
evolution of the binary prior to the explosion. All of these issues
play an important role in the final outcomes of NS kicks and are addressed
in the following. 

We note that the large body of literature of NS binaries and their
outcomes prohibits us from a detailed review of all previous works,
and we therefore refer the reader to various review papers (e.g. \citealp{Kalogera2007,Abadie2010,Postnov2014}).
These studies typically aimed to calculate the production rate of
specific types of objects and/or transients; we emphasize that our
aim is not to reproduce these extensive studies, but rather
to focus on the novel aspects of the close encounters following NS
natal kicks.
It's nevertheless important to point out at least briefly some of
these works. In particular over the last two decades Kalogera, Belczybski,
Fryer and collaborators have began and continued a long-term study
of binary stellar evolution (e.g. \citealt{Fryer1999,Kalogera2007}),
mostly using the STARTRACK population synthesis module they developed
to explore the rate of properties of a wide range of transients and
objects, and in particular GW sources and short-GRBs. Similar efforts
have been done by other groups, using other models such as the Scenario
Machine (e.g. \citealt{Lipunov2014} and references therein) and various
other models (e.g. \citealt{deMink2015} and references therein).

This paper is organized as follows: in Section \ref{sec:Calculating-the-minimal}
we present an analytical treatment to calculate the periapsis distance
of the binary post the SN accounting for a random position in the
initial Keplerian orbit, mass loss and a random kick velocity and
direction. In Section \ref{sub:Numerical-setup} we describe the population synthesis calculation. In Section \ref{sec:Results} we present the simulation
results for $587,019$ binary systems. In the following Section \ref{sec:Analytical-understanding}
we explain the results analytically. The implication of the results
are described in Section \ref{sec:Implications}, and the discussion
and summary of the paper is presented in Section \ref{sec:Discussion-and-Summary}.

\section{Calculating the minimal distance }
\label{sec:Calculating-the-minimal}
We begin by deriving the closest approach distance between the kicked
NS and the secondary. The closest approach is determined by the state
of the binary system right before the SN explosion, the prompt mass
lost during the SN, and the natal kick velocity vector. The binary
state before the SN is defined by the binary semi-major axis (sma),
$a$, the eccentricity $e$, the secondary mass, $m_{{\rm s}}$, the
primary mass before the SN, $m_{{\rm p}}$, and the specific orbital
phase of the stars in their orbit, given by the true anomaly, $\nu.$

The known solution to the two body Kepler problem is 
\begin{equation}
\frac{1}{r}=\frac{\mu Gm_{{\rm p}}m_{{\rm s}}}{l^{2}}\left(1+\sqrt{1+\frac{2{\rm E}l^{2}}{\mu G^{2}m_{{\rm p}}^{2}m_{{\rm s}}^{2}}\cos\nu}\right)
\end{equation}
where $r$ is the separation between $m_{{\rm p}}$ and $m_{{\rm s}}$,
$G$ is the Newton's constant, $\mu=m_{{\rm p}}m_{{\rm s}}/\left(m_{{\rm p}}+m_{{\rm s}}\right)$
is the reduced mass of the binary, ${\rm E}=\left(1/2\right)\mu\dot{r}^{2}-G\mu\left(m_{{\rm p}}+m_{{\rm s}}\right)/r$
it the total binary energy in the center of mass reference frame,
$l=\mu\left|\mathbf{r}\times\mathbf{\dot{r}}\right|$ is the system
total angular momentum and $\nu$ is the true anomaly.

The timescale of the SN explosion and the mass-loss is short compared
with the dynamical time of the system, and we can therefore assume that
during the SN the primary object is not changing its position, namely
$\mathbf{r}$ is not changed. After the SN the primary star with mass
$m_{p}$ undergoes prompt mass loss, and the NS remnant, with mass
$m_{{\rm NS}}$, is given a natal kick, $\Delta\mathbf{v_{{\rm kick}}}$.
In order to calculate the relative velocities between the binary components
just after the kick one needs to add the random natal kick velocity
vector to to the relative velocity between the objects, i.e. 

\begin{equation}
\mathbf{\dot{r}}_{{\rm pSN}}=\mathbf{\dot{r}}+\Delta\mathbf{v_{{\rm kick}}}\label{eq:relavtive v}
\end{equation}
 the subindex corresponds to post SN. Note that both $\mathbf{r}$
and its time derivative $\mathbf{\dot{r}}$ do not change under any
translation or boost transformation, due to the change in the center
of mass of the system.

As a result, the reduced mass $\mu_{{\rm pSN}}$, the total energy
${\rm E}_{{\rm pSN}}$ and the total angular momentum $l_{{\rm pSN}}$
change accordingly
\begin{equation}
\mu_{{\rm pSN}}=\frac{m_{{\rm NS}}m_{{\rm s}}}{m_{{\rm NS}}+m_{{\rm s}}}
\end{equation}
 
\begin{equation}
{\rm E}_{{\rm pSN}}=\frac{1}{2}\mu_{{\rm pSN}}\dot{r}_{{\rm pSN}}^{2}-\frac{G\mu_{{\rm pSN}}\left(m_{{\rm NS}}+m_{{\rm s}}\right)}{r_{{\rm pSN}}}
\end{equation}
\begin{equation}
l_{{\rm pSN}}=\mu_{{\rm pSN}}\left|\mathbf{r}\times\mathbf{\dot{r}}\right|.
\end{equation}
The solution for the new two body problem is then
\begin{equation}
\frac{1}{r}=\frac{\mu_{{\rm pSN}}Gm_{{\rm pSN}}m_{{\rm s}}}{l_{{\rm pSN}}^{2}}\left(1+\sqrt{1+\frac{2{\rm E}_{{\rm pSN}}l_{{\rm pSN}}^{2}}{\mu_{{\rm pSN}}G^{2}m_{{\rm NS}}^{2}m_{{\rm s}}^{2}}\cos\nu}\right).
\end{equation}
From the last equation we can easily compute the closest approach
of the NS to its companion, $r_{{\rm min}}$. If the binary is disrupted
i.e. ${\rm E_{pSN}}>0$ and $\mathbf{r}\cdot\mathbf{\dot{r}}>0$,
the closest separation is the separation at the moment of the SN,
otherwise it is given by 
\begin{equation}
\frac{1}{r_{{\rm min}}}=\frac{\mu_{{\rm pSN}}Gm_{{\rm pSN}}m_{s}}{l_{{\rm pSN}}^{2}}\left(1+\sqrt{1+\frac{2{\rm E}_{{\rm pSN}}l_{{\rm pSN}}^{2}}{\mu_{{\rm pSN}}G^{2}m_{{\rm NS}}^{2}m_{{\rm s}}^{2}}}\right).\label{eq:minimal r}
\end{equation}

\section{Population synthesis and numerical setup}
\label{sub:Numerical-setup}
In order to obtain the statistics of a large population we used the
publicly available open code - BSE \citep{Hurley2002}. The parameters
used are presented in Table \ref{tab:BSE parameters}. The initial
conditions for the binary population were constructed as to follow
the observed binary properties as reviewed by \citet{Duchene2013}.
We modeled $n=587,019$ binaries with primary mass $m_{{\rm p}}>8M_{\odot}$
from a Salpeter initial mass function, the sma, $a$, was randomly
chosen from a peak+power law distribution (see Figure 2 in \citealt{Duchene2013}). The mass ratio $q$ was randomly generated from a uniform
distribution for $a<0.45{\rm AU}$ (corresponding to orbital periods
of $<20\,$days for a $30M_{\odot}$ binaries); for wider separations
($0.45{\rm AU\leq}a\leq50{\rm AU}$) a power law distribution $f\left(q\right)\propto q^{-\gamma}$with
$\gamma=-2$ was used \citep{Moe2016} . The eccentricities of the
binaries were chosen from a thermal distribution $f\left(e\right)\propto e^{\eta}$
with $\eta=1$. The metallicity was chosen to be solar, namely $z=0.02$. 

We evolved the system until one of the companions becomes a NS through
a SN or accretion induced collapse (AIC). The population synthesis
code provides the orbital and stellar parameters, i.e. the sma $a$,
the eccentricity $e$, the mass and radius of the companion and the
mass of the NS. Next, we generated a random position in the binary
orbit by randomizing the mean motion and calculating the eccentric
anomaly, $E$ from it. The eccentric anomaly translates to separation,
$r$, by $r=a\left(1-e\cos E\right)$, and hence specifies the velocity
$\mathbf{v}$. In order to produce the natal kick velocity vector
we randomized the kick velocities chosen from a Maxwellian distribution
with a velocity dispersion of $\sigma=270{\rm kms^{-1}}$ \citep{Hobbs2005},
and isotropic unit vector around the NS. We used eq. (\ref{eq:minimal r})
to determine the minimal separation between the newly born NS and
its companion. Accounting for the closest approach together with the
radius of the companion at that point in the evolution we can determine
the outcome, and identify cases that produce direct collisions, $\mu$TDEs
or tidal captures. 

We flag a direct collision if 
\begin{equation}
r_{{\rm min}}<R_{{\rm s}}+R_{{\rm NS}}\label{eq:collision_criteria}
\end{equation}
where $R_{{\rm s}}$ is the secondary radius and $R_{{\rm NS}}$ is
the NS radius given from the population synthesis code. The criteria
for $\mu$TDE is given by the following condition 
\begin{equation}
R_{{\rm s}}+R_{{\rm NS}}<r_{{\rm min}}<R_{{\rm Tidal}}\label{eq:TDE_criteria}
\end{equation}
where $R_{{\rm Tidal}}=R_{{\rm s}}\left(2m_{{\rm NS}}/m_{{\rm s}}\right)^{1/3}$.
The criteria for tidal capture is 
\begin{equation}
r_{{\rm min}}<R_{{\rm s}}\left(\frac{Gm_{{\rm s}}}{R_{{\rm s}}v_{{\rm kick}}^{2}}\frac{m_{{\rm NS}}\left(m_{{\rm NS}}+m_{{\rm s}}\right)}{m_{{\rm s}}^{2}}\right)^{1/6}\label{eq:Tidal_Capture_Criteria}
\end{equation}
is taken from \citet{Fabian1975}.

Note that when $R_{{\rm Tidal}}<R_{s}$, the tidal radius is inside
the companion radius. In such cases we flag the outcome as a collision.

\begin{table*}
\begin{tabular}{|c|c|c|c|c|c|c|}
\hline 
\multicolumn{7}{|c|}{BSE parameters}\tabularnewline
\hline 
\hline 
neta=0.5 & alpha1=1 & ftlag=1 & bhflag=0 & pts1=0.05 & beta=0.125 & epsnov=0.001\tabularnewline
\hline 
bwind=0 & lambda=0.5 & ifflag=0 & nsflag=1 & pts2=0.01 & xi=1 & eddfac=1\tabularnewline
\hline 
hewind=1 & ceflag=0 & wdflag=1 & mxns=3 & pts3=0.02 & acc2=1.5 & gamma=-1\tabularnewline
\hline 
\end{tabular}

\protect\caption{\label{tab:BSE parameters}The parameters used for the population
synthesis. For detail explanation see \citet{Hurley2002}}
\end{table*}

\section{Results}
\label{sec:Results}

\subsection{Stellar/dynamical evolution before the SN}

In this subsection we present the evolution of a population binaries
from the initial conditions at zero age main sequence (ZAMS) until
the formation of the first NS in the system, either through a core-collapse
SN or an AIC, if such occurs, as determined by the BSE code \citep{Hurley2002}.
These set the conditions of the binaries just before the SN kick.

\begin{figure*}
\includegraphics[width=17cm]{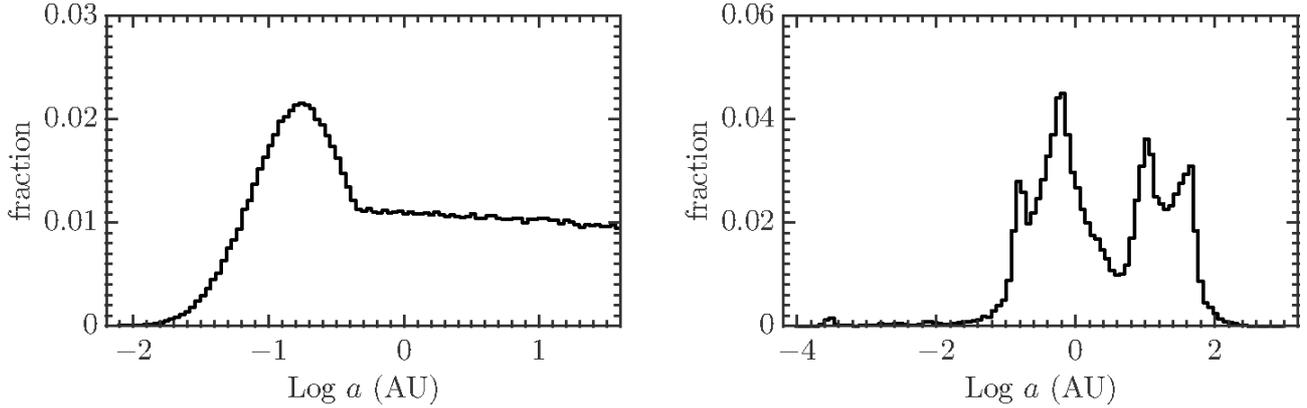}\protect\caption{\label{fig:sma_dis_before}\textbf{Left panel:} semi-major axis distribution
of the entire sample, $587,019$ systems, at zero age main-sequence.
\textbf{Right panel:} semi-major axis, $a_{{\rm pre}}$, distribution
of the surviving binaries, $144,915$ systems, after stellar evolution
just before the SN (not necessarily at the same time). Due to stellar
evolution about three quarters the binaries either disrupted or merged. }
\end{figure*}

In the left panel of Figure \ref{fig:sma_dis_before} we present the
initial sma, $a$, distribution at ZAMS, the beginning of the integration,
for the $587,019$ systems. On the right panel we present the sma
distribution just before the SN, $a_{{\rm pre}}$. The number of systems that
survive as binaries up to the stage where a SN explodes or an AIC event takes place leaving behing 
a NS is $144,915$. About three quarters of the systems
the binary components merged or the binary was disrupted before any
NS was formed. The remaining systems have a wider sma distribution
compared with the initial population due binary interaction or stellar
evolution, e.g. common envelope, tidal friction or mass loss from
one of the companions.

\begin{figure}
\includegraphics[width=8cm]{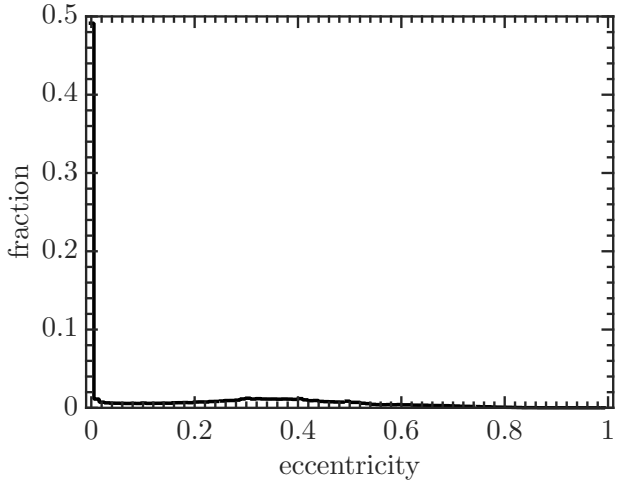}

\protect\caption{\label{fig:eccentricity}Eccentricity distribution of the surviving
binaries ($144,915$ systems). Approximately $55\%$ of the binaries
are circular due to tidal interactions.}
 
\end{figure}

In Figure \ref{fig:eccentricity} we present the eccentricity distribution
of the population just before the SN. One can see the effects of binary
evolution, mainly the circularization of the binaries due to tidal
interaction during the giant phase.

The post-SN outcome strongly depends on the binary mass and separation.
Figure \ref{fig:Left-panel:average_mass_massloss} shows the average
mass of binaries just prior to the SN as a function of their orbital
separation. Binaries with separation wider than $\sim10{\rm AU}$
have a typical mass of $m_{{\rm b}}\approx15M_{\odot}$ while binaries
with separation of $\sim1{\rm AU}$ have a wider mass range between
$m_{{\rm b}}\approx20-28M_{\odot}$. Closer binaries have lower total
masses ranging between $m_{{\rm b}}\approx2-10M_{\odot}.$ In the
right panel of Figure \ref{fig:Left-panel:average_mass_massloss}
we plot the averaged normalized binary mass loss, $\left\langle \Delta m/m_{{\rm b}}\right\rangle $,
namely the ratio between the mass loss due to the SN and the binary
total mass prior the SN. \citet{Hills1983} showed that the normalized
mass loss is a measure for dissociation. All binaries dissociate if
the following condition is met 
\begin{equation}
\frac{\Delta m}{m_{{\rm b}}}\geq\frac{r}{2a_{{\rm pre}}}\left(1-\left(\frac{\Delta v_{{\rm kick}}}{v_{{\rm c}}}\right)^{2}-2\left(\frac{v}{v_{{\rm c}}}\right)\left(\frac{\Delta v_{{\rm kick}}}{v_{{\rm c}}}\right)\cos\theta\right)
\end{equation}
where $v_{c}$ is the relative orbital velocity between the stars
when $r=a_{{\rm pre}}$, $v$ is the relative orbital velocity at
the moment of the explosion and $\theta$ is the angle between $\mathbf{v}$
and $\mathbf{\Delta v_{{\rm kick}}}$. Hence for sufficiently large
mass loss any binary would dissociate.

\subsection{Post-SN dynamical evolution}

\begin{figure*}
\includegraphics[width=17cm]{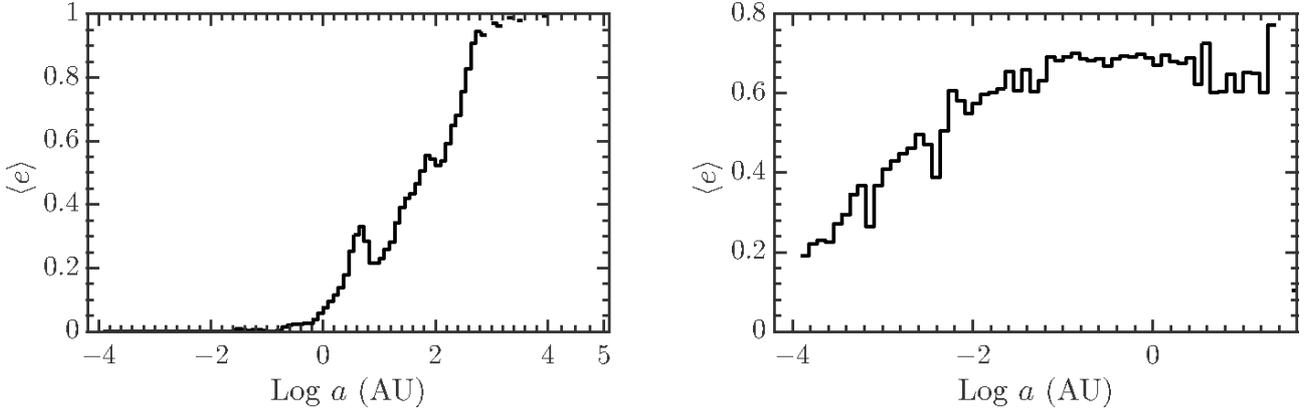}\protect\caption{\label{fig:Averaged-eccentricity-of}\textbf{Left panel:} The average
eccentricity of the binary prior to the SN. All systems with separation
of $<1\ {\rm AU}$ circularized due to tidal interactions during the
course of the binary stellar evolution. \textbf{Right panel: }The
average eccentricity of bound binaries after the SN and natal kick. }
 
\end{figure*}

\begin{figure}
\includegraphics[width=8cm]{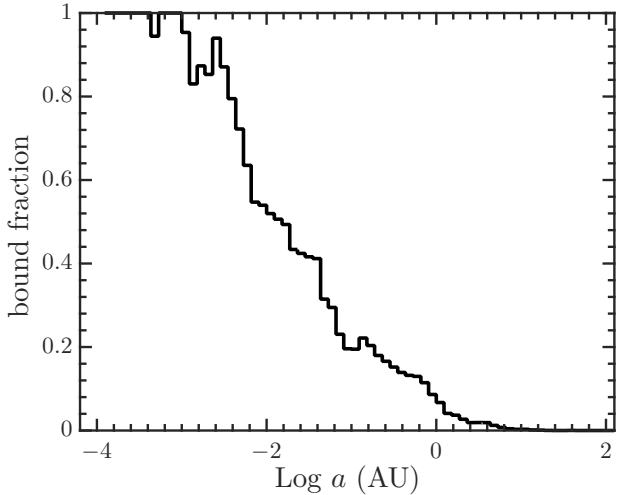}\protect\caption{\label{fig:Fraction-of-bound}The fraction of bound systems as a function
of semi major axis. An inverse trend is visible. Roughly all systems
are disrupted at $\sim1\ {\rm AU}.$}
\end{figure}

Figure \ref{fig:Averaged-eccentricity-of} shows the average eccentricity
as a function of separation before the SN (left panel) and immediately
after the SN (right panel). In the left panel we explicitly see the
circularization of the binaries with separation smaller than $\sim1\ {\rm AU}$
and a monotonically increasing trend with increasing separation. Disrupted
binaries have a hyperbolic trajectory that can result in two outcomes:
(\mbox{i}) the NS is kicked away from its companion never to interact
with it (\mbox{ii}) the NS is kicked towards its companion and may
interact with it as discussed in the previous subsection. In Figure
\ref{fig:Fraction-of-bound} we present the fraction of post-kick
bound systems as a function of $a_{{\rm pre}}$. The data shows an
inverse trend of the fraction of bound systems with sma, as expected,
given the ratio of the kick velocity $v_{{\rm kick}}$ and the orbital
velocity, $v_{0}$ at the moment of the SN. The data indicates that
at separation of $\sim1{\rm AU}$ most binaries are disrupted following
the natal kick imparted by the SN explosion.


\subsubsection{The Distribution of closest approaches
after the natal kick}
\label{sub:Closest-approach-after}
\begin{figure}
\includegraphics[width=8cm]{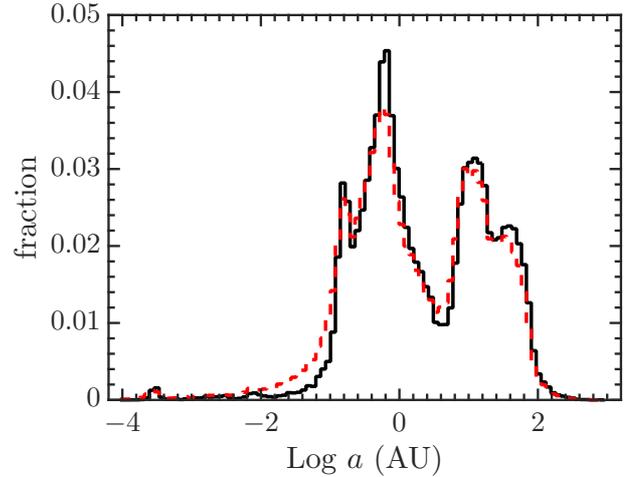}

\protect\caption{\label{fig:Rmin_Separation}Closest approach distribution compared
to the separation between the companions. Black solid line is the
separation distribution at the moment of NS formation. Red dashed
line is the distribution of the closest approach. It is clear that
for the lower end of the distribution (smaller distances) the fraction
of the closest approach is higher.}
\end{figure}

In this subsection we present the main numerical results; the distribution
of the closest approaches of the newly born NSs to their companions.
In Figure \ref{fig:Rmin_Separation} we compare the distribution of
the closest approaches with the \textit{separation} (and not the sma)
between the objects just before the SN. 

Most of the systems will not interact after the NS kick. As discussed
above, the systems which do interact can be divided between three
non-trivial outcomes: (\mbox{i}) Collision (eq. \ref{eq:collision_criteria})
(\mbox{ii}) $\mu$TDE (eq. \ref{eq:TDE_criteria}). (\mbox{iii}) Tidal
capture (eq. \ref{eq:Tidal_Capture_Criteria}). We note that the majority
of systems that result in one of these three outcomes are considered
bound in the moment following the SN, ${\rm E_{pSN}<0}$. Note that
the tidal capture case includes binaries that were bound following
the SN, and therefore tidal capture is somewhat of a misnomer. However,
the strong tidal interaction dissipates a significant fraction
of the orbital energy as to dynamically capture the binary into a
much closer binary compared with its face value properties post-kick
in the absence of tidal interactions. 

In Table \ref{tab:Summary-of-the-Results} we present a summary of
the results. The number of systems resulting in a direct collision
is $2,176$ which is $\sim0.9\%$ of the surviving binaries and $\sim0.37\%$
of the total sample. In order to provide rate estimate per galaxy,
we consider fiducial number for the Milky-Way galaxy, taking $2\times10^{11}$ stars, of which $4\times10^{8}$
are stars with mass greater than $8M_{\odot}$. Taking these values, and assuming the stars in the Galaxy
were formed continuously over $10\rm Gyr$, our results translate to a post-SN close encounters rate of $\approx1\times10^{-4}\rm yr^{-1}$ Among
thes systems $1,809$ $\left(\sim83.1\%\right)$ were bound the moment following the SN
and $367$ $\left(\sim26.9\%\right)$ were unbound. The number of
systems that satisfied the $\mu$TDE criteria is $309$ (giving a $\mu$TDE rate of ,$\sim2\times10^{-5}{\rm yr^{-1}}$)
of which $236$ $\left(\sim76.3\%\right)$ were bound and $73$ $\left(\sim23.7\%\right)$
were unbound. The number of tidally captured system is $156$ (production rate
of $\sim1\times10^{-5}{\rm yr^{-1}}$), comprised out of $151$ $\left(\sim96.7\%\right)$
bound and $5$ $\left(\sim3.3\%\right)$ unbound systems.

\begin{figure}
\includegraphics[width=8cm]{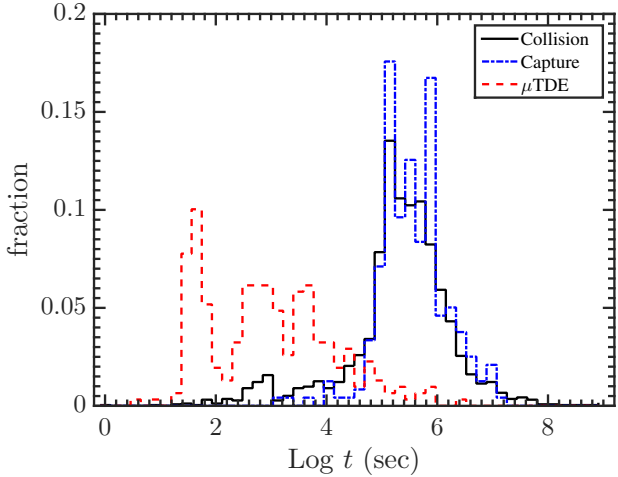}\protect\caption{\label{fig:Elapsed-time-distribution} The delay time distribution
between the SN explosion and the periapsis passage for close encounters
of NSs and their companion stars. of the NS near its companion Black
solid line depicts direct collisions, blue dash point line depicts
tidal captures and red dashed line depicts $\mu$TDEs.}
\end{figure}

\begin{figure*}
\includegraphics[width=17cm]{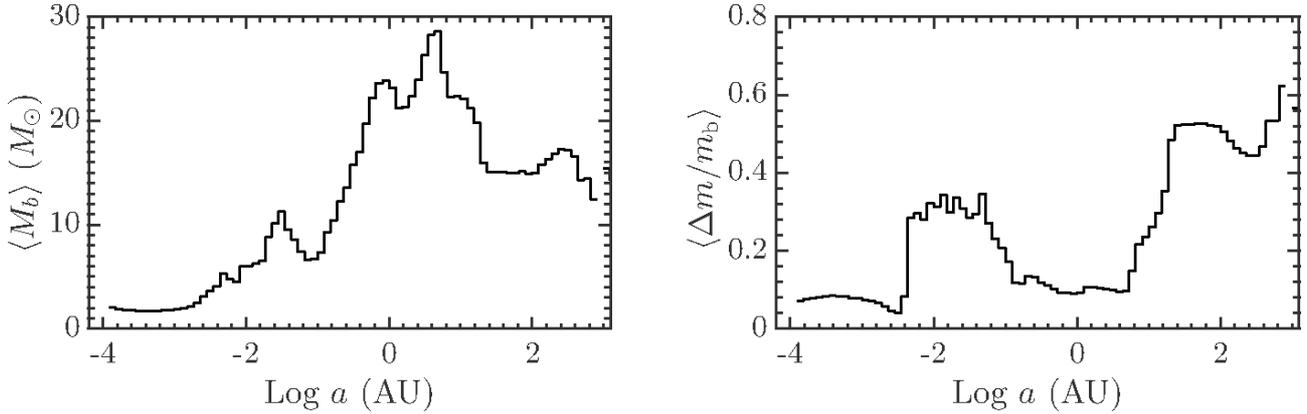}\protect\caption{\textbf{\label{fig:Left-panel:average_mass_massloss}Left panel:}
The average binary mass, $\left\langle m_{{\rm b}}\right\rangle $
just prior to the SN, as a function of the orbital separation. \textbf{Right
panel:} The average of the normalized mass loss $\left\langle \Delta m/m_{{\rm b}}\right\rangle $,
as a function of orbital separation. The higher the value $\Delta m/m_{{\rm b}}$
, the higher is the probability to disrupt the binary. }
\end{figure*}

The close encounters occur when the stars in the binary system are
sufficiently close, namely at periapsis of the trajectory (either
elliptical or hyperbolic). The elapsed time since the SN down to the
periapsis approach is given by the known formula from Kepler equations.
For a bound binary,
\[
\Delta t=\sqrt{\frac{a_{{\rm post}}^{3}}{GM_{{\rm binary}}}}\left(\frac{\pi}{2}-\arcsin\left(\frac{1-\tilde{a}}{e}\right)\right)\quad{\rm if}\ \mathbf{r}\cdot\mathbf{\dot{r}}<0
\]

\begin{equation}
\Delta t=\sqrt{\frac{a_{{\rm post}}^{3}}{GM_{{\rm binary}}}}\left(2\pi-\left(\frac{\pi}{2}-\arcsin\left(\frac{1-\tilde{a}}{e}\right)\right)\right)\quad{\rm if}\ \mathbf{r}\cdot\mathbf{\dot{r}}>0\label{eq:Delay_time_bound}
\end{equation}
and for an unbound binary, 
\begin{equation}
\Delta t=\sqrt{\frac{-a^{3}}{GM_{{\rm binary}}}}\log\left(1-\tilde{a}+\sqrt{\tilde{a}^{2}-2\tilde{a}-\left(e^{2}-1\right)}\right),
\end{equation}
where $G$ is Newton's constant; $\tilde{a}\equiv r/a$ is the ratio
between the separation and the sma immediately after the SN; and $M_{{\rm binary}}$
and $e_{{\rm post}}$ are the post-SN binary mass and eccentricity,
respectively. Figure \ref{fig:Elapsed-time-distribution} presents
the distribution of the elapsed times since the SN (``delay times''
hereafter). The distributions are shown for each type of interaction.
The delay time distribution for collisions and tidal-captures follow
similar behavior, with $\left\langle \Delta t\right\rangle _{{\rm collision}}\approx4.7\times10^{6}{\rm }$
sec and $\left\langle \Delta t\right\rangle _{{\rm capture}}\approx9.3\times10^{5}{\rm }$sec.
The $\mu$TDEs follow a different distribution with much shorter delay
times, with an average time of $\left\langle \Delta t\right\rangle _{\mu{\rm TDE}}\approx3\times10^{4}{\rm sec}\approx8$
hours. 

\begin{table*}
\begin{tabular*}{1\columnwidth}{@{\extracolsep{\fill}}ccccc}
\hline 
outcome & Systems & Rates $\left({\rm yr^{-1}}\right)$ & Bound systems  & Unbound systems \tabularnewline
\hline 
\hline 
Collisions & $2176\ \left(0.37\%\right)$ & $1\times10^{-4}$ & $1809\ \left(83.1\%\right)$ & $367\ \left(16.9\%\right)$\tabularnewline
\hline 
$\mu$TDE & $309\ \left(0.05\%\right)$ & $2\times10^{-5}$ & $236\ \left(76.3\%\right)$ & $73\ \left(23.7\%\right)$\tabularnewline
\hline 
Tidal capture & $156\ \left(0.02\%\right)$ & $1\times10^{-5}$ & $151\ \left(96.7\%\right)$ & $5\ \left(3.3\%\right)$\tabularnewline
\hline 
\end{tabular*}\protect\caption{\label{tab:Summary-of-the-Results}Summary of the possible close encounter
outcomes from NS natal kick (and the rates per Milky-Way galaxy) .The
numbers are out of a total of $587,019$ modeled systems. }
 
\end{table*}

\subsubsection{Dependence of post-SN outcomes on pre-SN binary configurations and
kicks}

In Figure \ref{fig:sma-distribution-of} we present the pre-explosion
sma distribution for each type of interactions Direct collisions and
tidal captures follow similar distributions with an sma range of $\sim0.05-1{\rm AU}$
with $\left\langle a_{{\rm pre}}\right\rangle _{{\rm collision}}\approx0.44$
AU and $\left\langle a_{{\rm pre}}\right\rangle _{{\rm capture}}\approx0.41{\rm }$AU,
while $\mu$TDEs follow a much tighter sma of $10^{-4}-10^{-1}{\rm }$
AU with $\left\langle a_{{\rm pre}}\right\rangle _{\mu{\rm TDE}}\approx0.012{\rm }$
AU. These follow the same trends as
seen for the delay-time distributions (Figure \ref{fig:Elapsed-time-distribution}),
as expected from the relation between delay times and pre-SN separations.

Figure \ref{fig:Kick_Dist} shows the kick velocity distribution for
all types of systems found to be strongly interacting. The averaged
velocity distribution for all types of systems is $\sim349{\rm kms^{-1}}$
(compared with $\sim430{\rm kms^{-1}}$ for the primordial averaged
kick, taken from the Maxwellian distribution; not shown) and the standard
deviation (std) is $\sim151{\rm kms^{-1}}$ (compared with $\sim180{\rm kms^{-1}}$for
the Maxwellian distribution). The systems undergoing collisions have
a kick distribution with an average value of $v_{{\rm k}}\approx345{\rm kms^{-1}}$
with std of $\sim140{\rm kms^{-1}}$. Tidal captured systems experience
an average kick velocity of $v_{{\rm k}}\approx223{\rm kms^{-1}}$
with std of $\sim88{\rm kms^{-1}}.$ The $\mu$TDEs have an average
kick velocity of $v_{{\rm k}}\approx439{\rm kms^{-1}}$ with std of
$\sim188{\rm kms^{-1}.}$ These result are also consistent with Figure
\ref{fig:sma-distribution-of} indicating that shorter period binaries
(higher orbital velocities) are the progenitors for $\mu$TDEs.

\subsubsection{Distribution of stellar components}

The final outcome of the post-SN close encounters is strongly dependent
on the radius of the stellar companion of the NS, which itself is
related to the mass and stellar evolutionary stage of the star. These
dependencies are well reflected in the distributions of these properties,
as we show below.

The distribution of companion radii is shown in Figure \ref{fig:Radius_Companion}.
The average values of the companion radius in the collision, capture
and $\mu$TDE cases are $\left\langle R_{{\rm companion}}\right\rangle _{{\rm collision}}\approx5.17R_{\odot}$,
$\left\langle R_{{\rm companion}}\right\rangle _{{\rm capture}}\approx3.92R_{\odot}$
and $\left\langle R_{{\rm companion}}\right\rangle _{{\rm \mu TDE}}\approx0.26R_{\odot}$,
respectively. 

The stellar types of the NS companions are presented in Figure \ref{fig:Companion-types},
where the stellar type classification are directly taken from the the
BSE classification scheme. The majority of collisions and tidal captures
are of MS companions; $\sim86.8\%$ for the collision case and $\sim94.9\%$
for the tidal capture case. The $\mu$TDE typical companion, however
are very different; $\sim44\%$ of the disrupted companions are Hertzsprung
gap He stars and $\sim29\%$ are stripped MS stars (He stars).

\begin{figure}
\includegraphics[width=8cm]{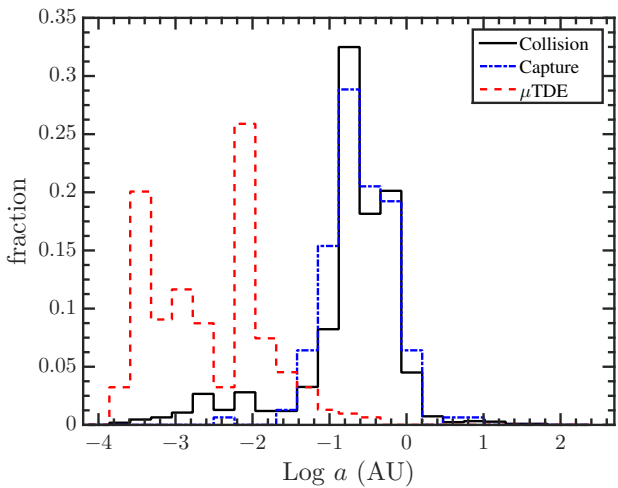}\protect\caption{\label{fig:sma-distribution-of}The pre-SN sma distribution of all
(post-SN) interacting binaries. Colors and line types the same as
in figure \ref{fig:Elapsed-time-distribution}}
\end{figure}

\begin{figure}
\includegraphics[width=8cm]{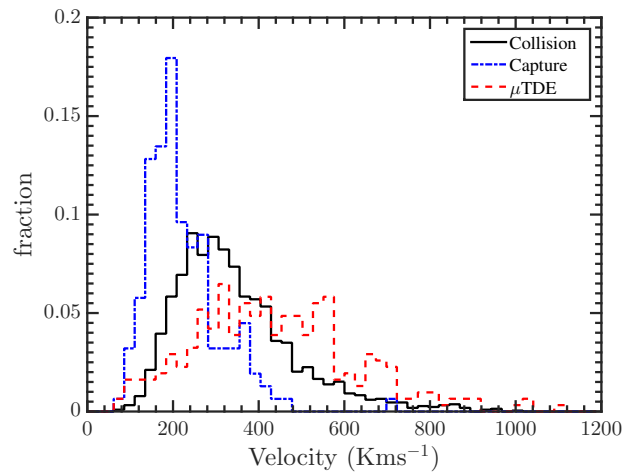}

\protect\caption{\textbf{\label{fig:Kick_Dist}} Kick velocity distribution all interacting
systems:Colors and line types the same as in figure \ref{fig:Elapsed-time-distribution} The
$\mu$TDE cases have a wider spread and higher values of the kick
velocity.}
\end{figure}

\begin{figure}
\includegraphics[width=8cm]{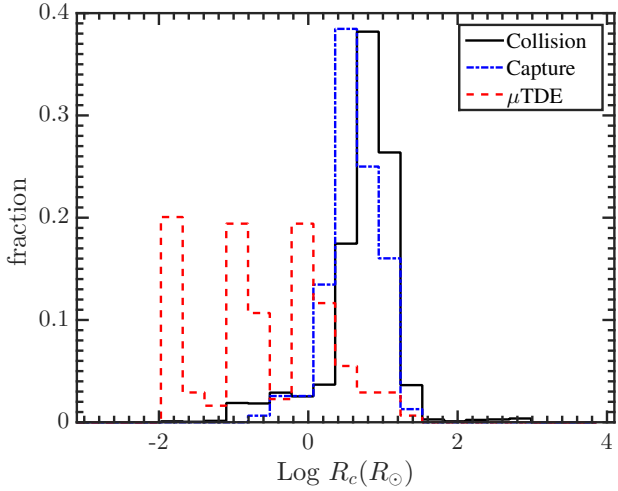}

\protect\caption{\label{fig:Radius_Companion}Distribution of the companion radii for
each type of interaction. Colors and line types the same as in figure
\ref{fig:Elapsed-time-distribution} Collisions and tidal captures
typically occur with large, $\sim10R_{\odot}$ stars while $\mu$TDEs
cases are typically disruptions of WD companions.\textbf{ }}
\end{figure}

\begin{figure*}
\includegraphics[width=8cm]{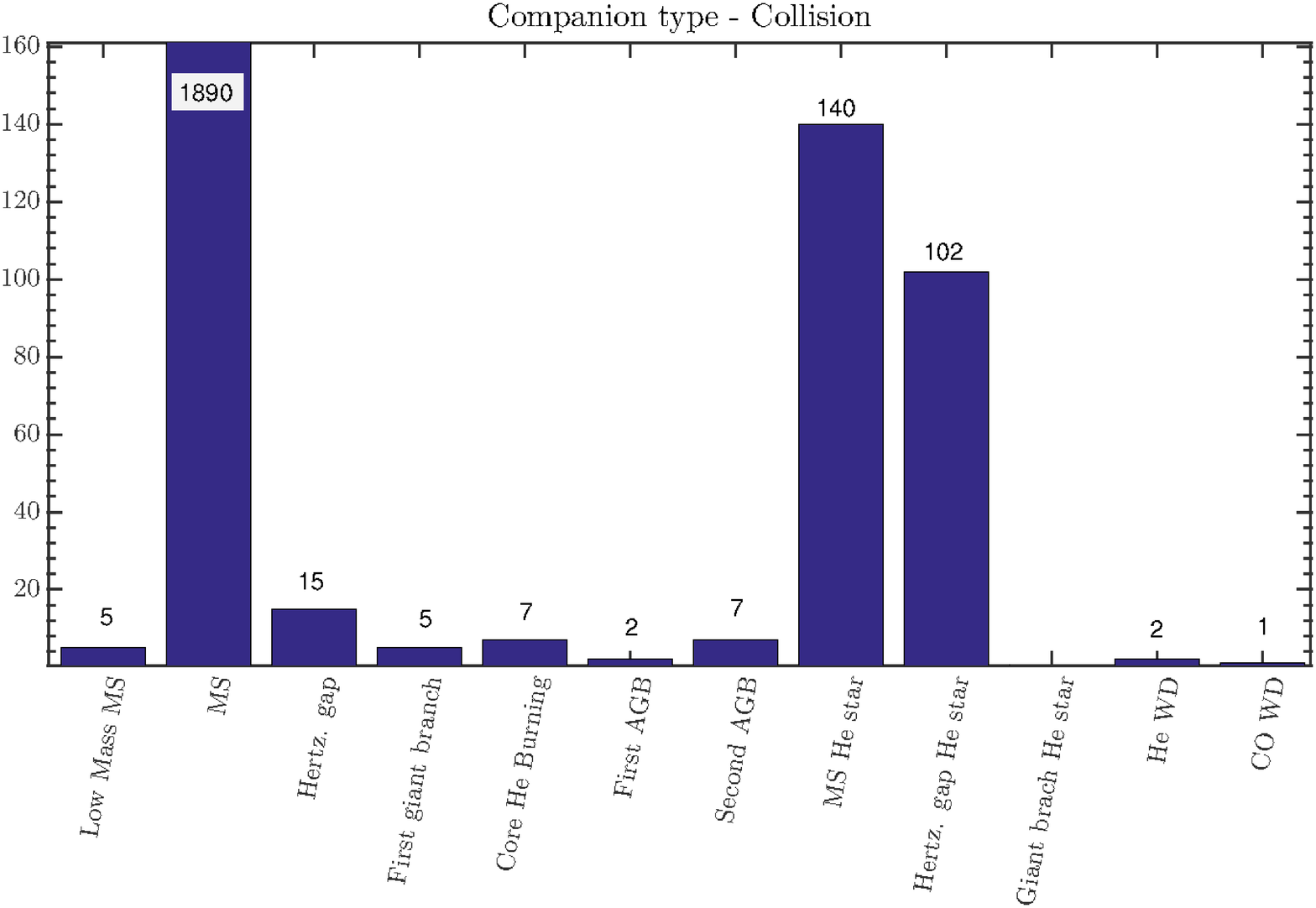}$\ $\includegraphics[width=8cm]{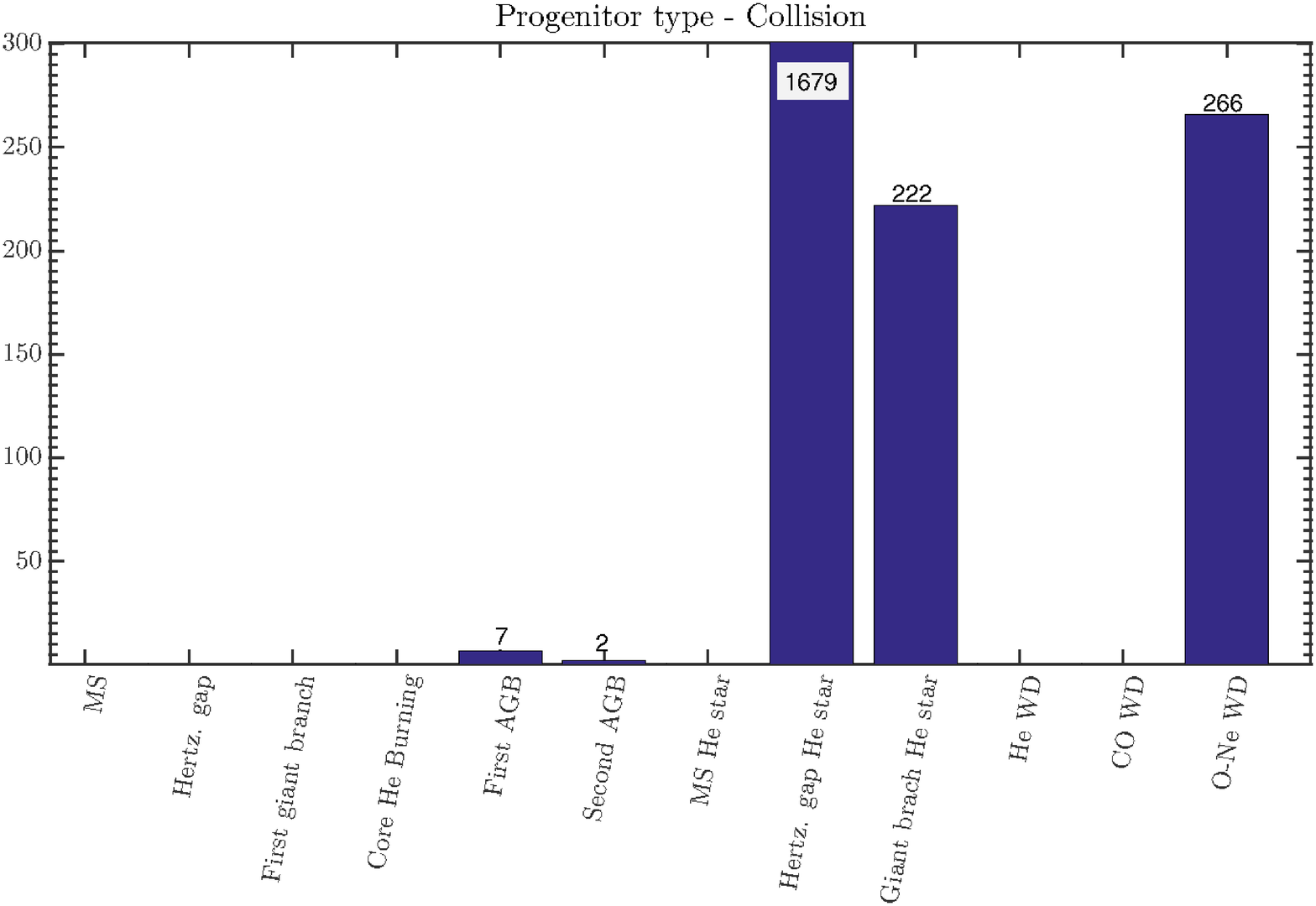}

\includegraphics[width=8cm]{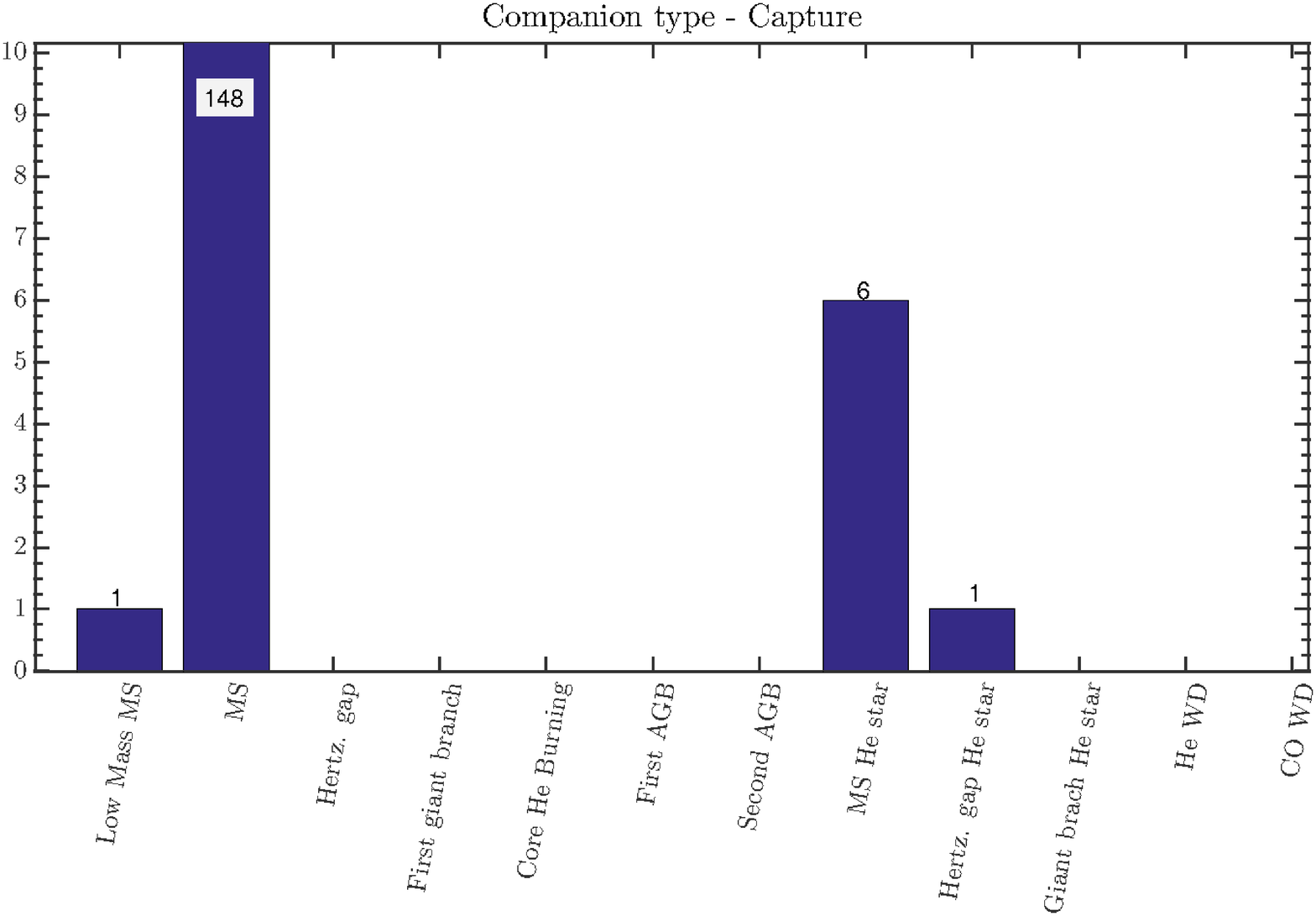}$\ $\includegraphics[width=8cm]{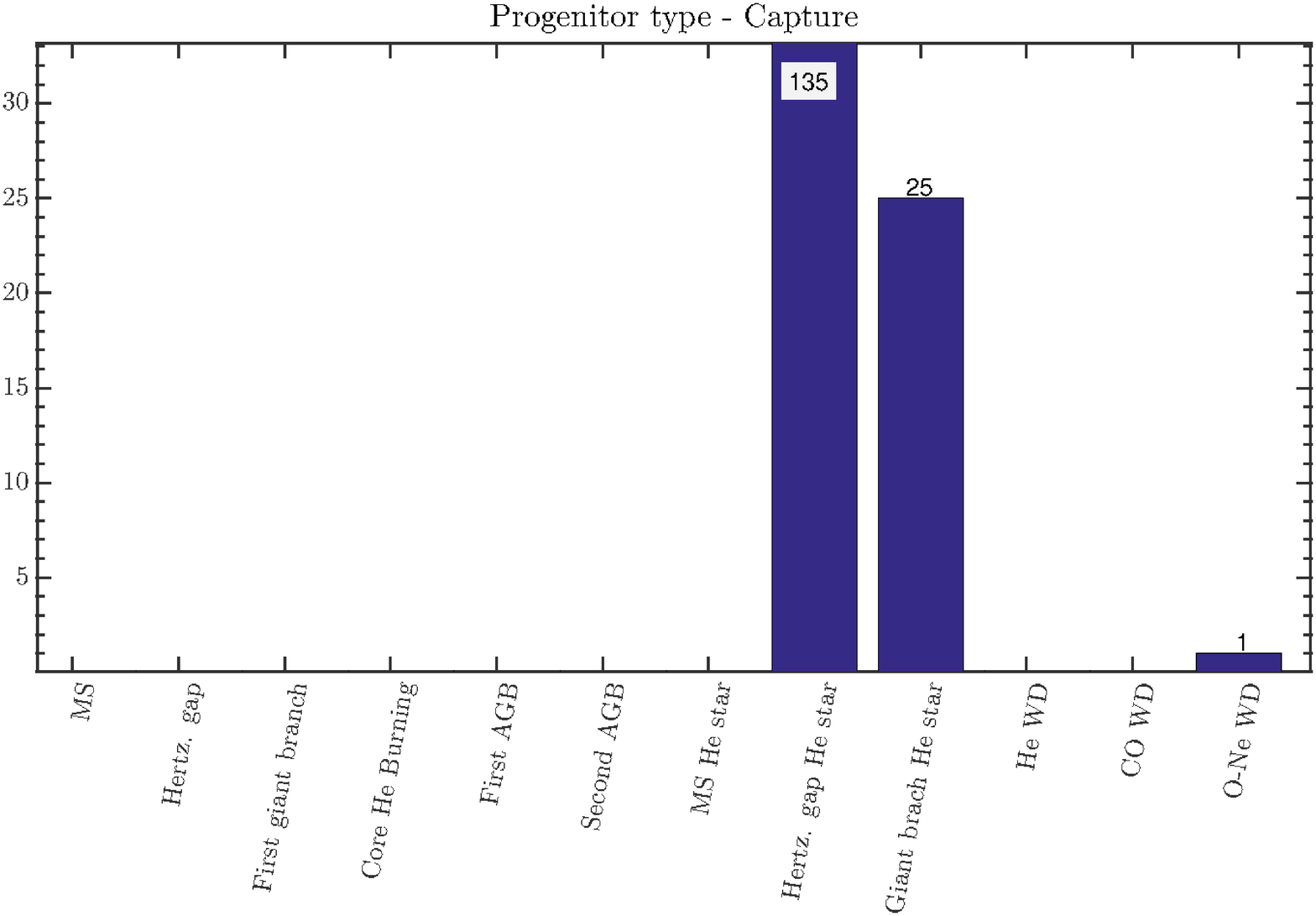}

\includegraphics[width=8cm]{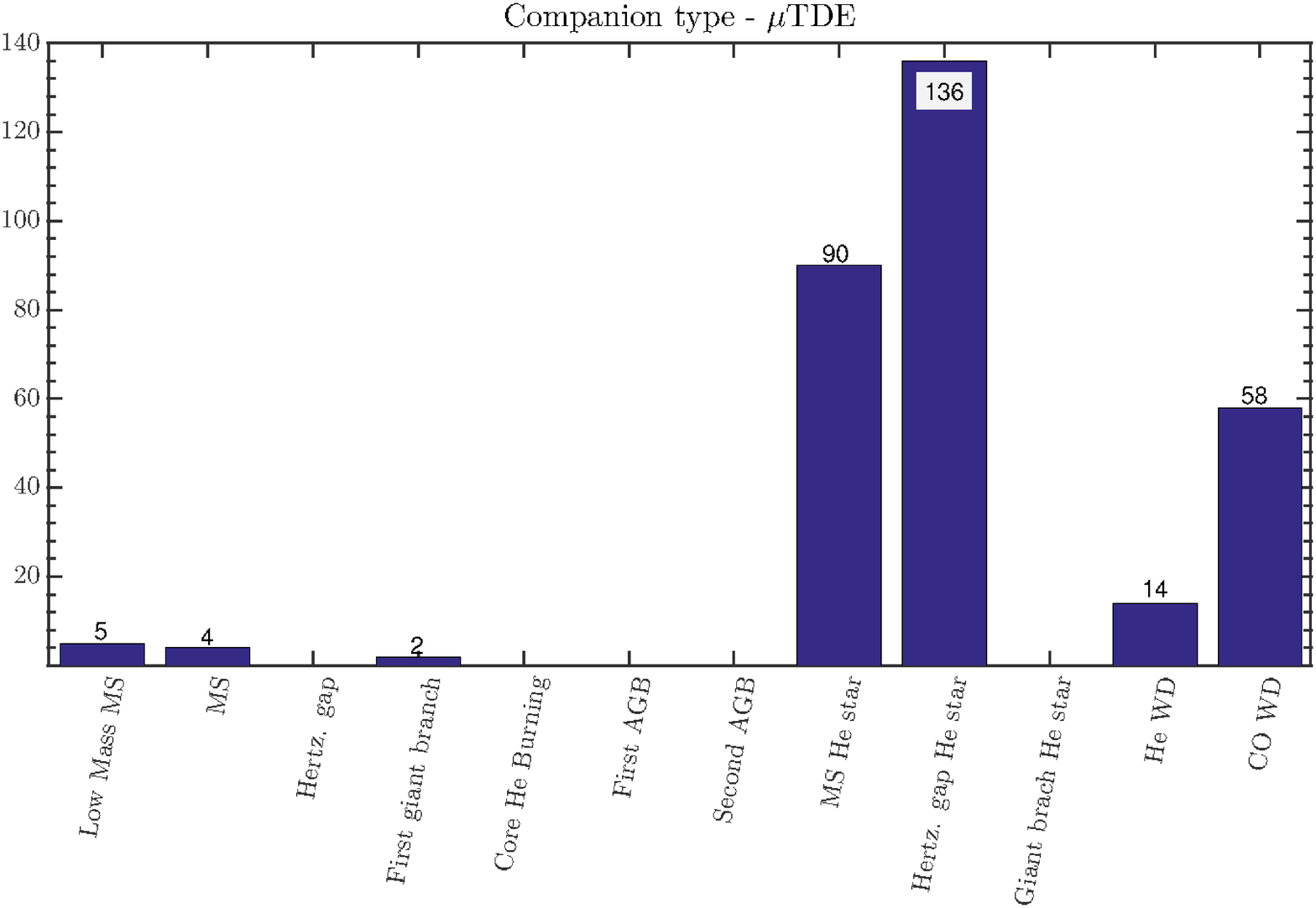}$\ $\includegraphics[width=8cm]{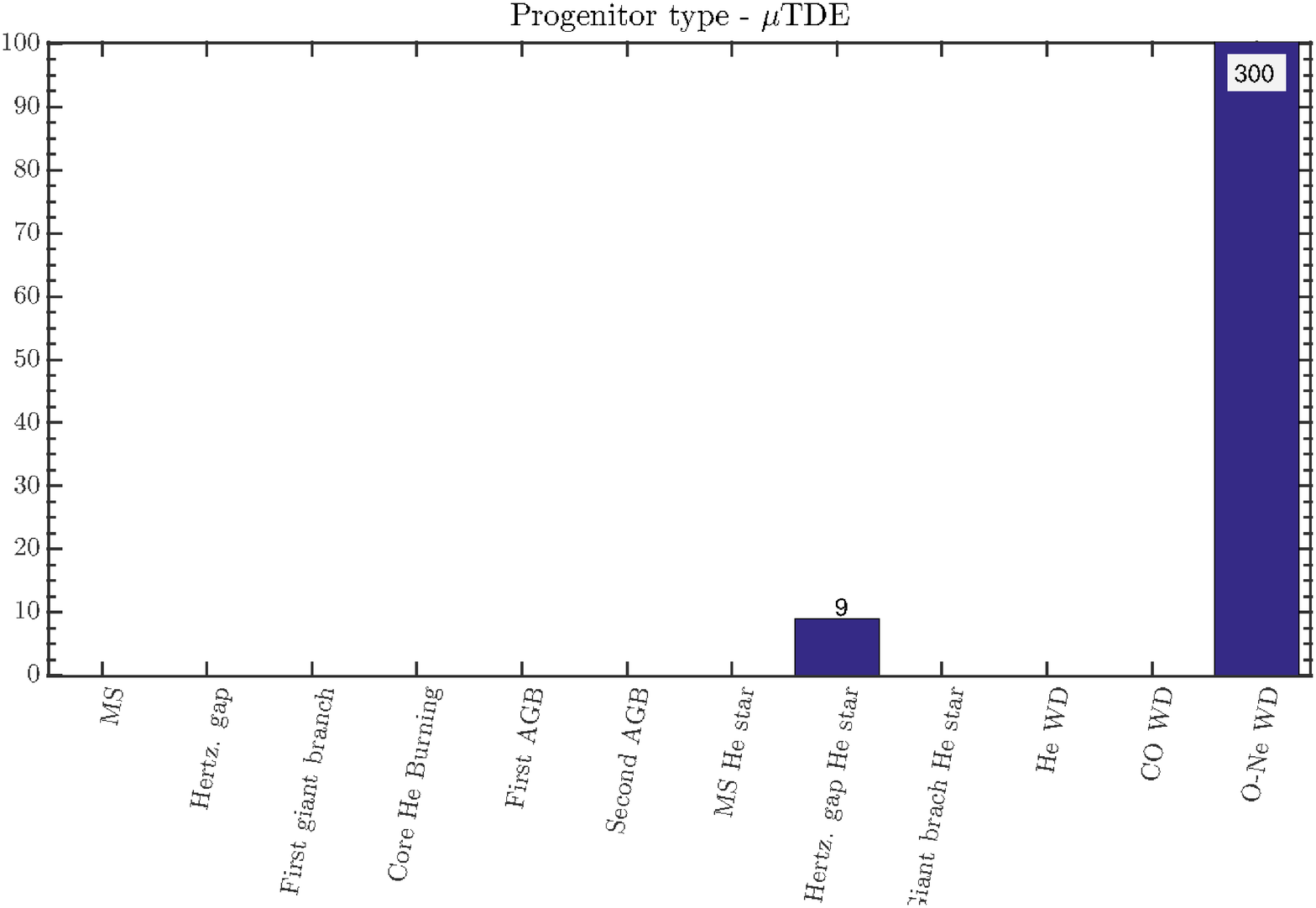}\protect\caption{\label{fig:Companion-types}Distribution of NS-companion and progenitor
stellar types. \textbf{Upper left panel:} Collisions, $\sim86.8\%$
of all collisions have a MS companion. \textbf{Middle left panel:}
Tidal capture, $\sim94.9\%$ of all tidal capture NSs in a new orbit
have a MS companion. \textbf{Bottom left panel:} $\mu$TDE, $\sim44\%$
of all $\mu$TDE have a Hertzsprung gap He star companions.\textbf{
Upper right panel:} Collisions, $\sim77\%$ of all collisions are
from a Hertzsprung gap He star progenitors. \textbf{Middle right panel:}
Tidal capture, $\sim86.5\%$ of progenitors are\textbf{ }from a Hertzsprung
gap He star.\textbf{ Bottom right panel:} $\mu$TDE, $\sim97\%$ of
progenitors are of O-He WD stellar type.}
\end{figure*}

\begin{figure}
\includegraphics[width=8cm]{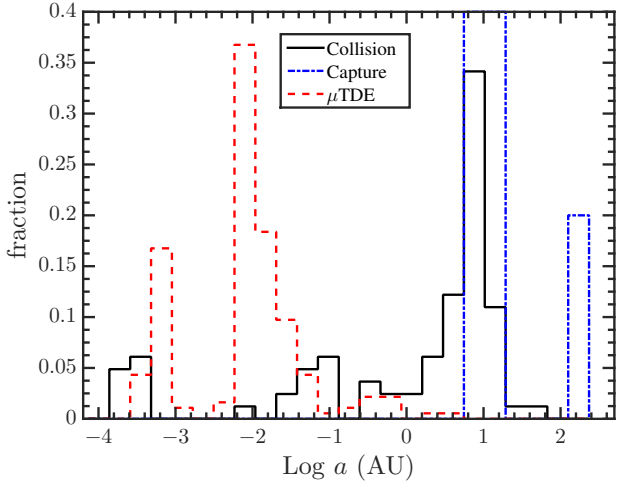}\protect\caption{\label{fig:AIC_sma-distribution-by}The pre-SN sma distribution of
all (post-SN) interacting binaries with AIC-formed NSs, with kick
velocity dispersion of $\sigma_{{\rm AIC}}=20{\rm kms^{-1}}$: Colors
and line types the same as in figure \ref{fig:Elapsed-time-distribution}
This distribution resembles the distribution in Figure \ref{fig:sma-distribution-of}
but with a shift for larger values of sma due to the smaller kick. }

\end{figure}

In order to understand the results one should also consider the origin
of the NSs involved. We find that the majority of the NS progenitors
in the collision and tidal capture cases are Hertzsprung gap He stars
( $\sim77.2\%$, $\sim86.5\%$, respectively), i.e. evolved massive
stars that will undergo core-collapse SN to produce a NS. However,
the vast majority of the progenitors of the $\mu$TDE are a O-Ne WD,
with $\sim97\%$ of all such events. This leads to the conclusion
that in these latter cases the NS were formed via accretion onto a
WD, resulting in an AIC that produces the NS. It was suggested that
NSs formed through AICs receive a much smaller natal kick than core-collapse
- formed NSs, therefore, we tried to account for this by changing
the velocity dispersion of the natal kick of all the O-Ne WD NS progenitors
in our runs to $\sigma_{{\rm AIC}}=20{\rm kms^{1}}$ ($90\%$ of the
kicks are lower than $50{\rm kms^{-1}}$, as suggested by \citep{Podsiadlowski2004}).
The results are presented in Table \ref{tab:AIC_table}. Figure \ref{fig:AIC_sma-distribution-by}
(like Figure \ref{fig:sma-distribution-of}) shows the distribution
of sma for all three close-encounter outcomes for the AIC events.
The data indicates that, as expected, the sma distribution for low
natal kicks is wider than the distribution for the original higher
kicks. 

\begin{table*}
\begin{tabular*}{1\columnwidth}{@{\extracolsep{\fill}}ccccc}
\hline 
outcome & Systems (from AIC events) & Rates $\left({\rm yr^{-1}}\right)$ & Bound systems  & Unbound systems\tabularnewline
\hline 
\hline 
Collisions & $82\ \left(3.8\%\right)$ & $5\times10^{-6}$ & $80\ \left(97.5\%\right)$ & $2\ \left(2.5\%\right)$\tabularnewline
\hline 
$\mu$TDE & $185\ \left(8.6\%\right)$ & $1\times10^{-5}$ & $185\ \left(100\%\right)$ & $0$\tabularnewline
\hline 
Tidal capture & $5\ \left(0.2\%\right)$ & $3\times10^{-7}$ & $5\ \left(100\%\right)$ & $0$\tabularnewline
\hline 
\end{tabular*}\protect\caption{\label{tab:AIC_table}Summary of the possible close encounter outcomes
from NS natal kick for AIC-formed NSs (and the rates per Milky-Way
galaxy). Natal kicks drawn from a Maxwellian distribution with $\sigma_{{\rm AIC}}=20{\rm kms^{-1}}.$ }
\end{table*}

\subsection{Comparison to different $\sigma_{{\rm kick}}$}
\label{sub:Comparison-to-different}
In the previous subsection we described the results for $\sigma_{{\rm kick}}=270{\rm kms^{-1}}$,
in this subsection we will compare the results to the same simulation
for a different natal kick velocity dispersion, $\sigma_{{\rm kick}}=190{\rm kms^{-1}}$
\citep{Hurley2002}, in order to explore the dependence of the outcomes
on the assumed distribution of natal kicks. The number of events in
all three outcomes is $2,560$ compared with $2,641$ events with
the higher velocity dispersion, this change is insignificant (smaller
than $0.02\%$). We found $2,108$ direct collisions, $271$ $\mu$TDEs
and $181$ tidal captures into a close binary. The sma distribution
(not shown) is somewhat altered (compared with Figure \ref{fig:sma-distribution-of}),
and gives rise to larger separations, on average, as expected from
a smaller natal kick. The average values for the different type of
interaction are $\left\langle a_{{\rm pre}}\right\rangle _{{\rm collision}}\approx0.5{\rm AU}$
, $\left\langle a_{{\rm pre}}\right\rangle _{{\rm capture}}\approx0.47{\rm AU}$
and $\left\langle a_{{\rm pre}}\right\rangle _{\mu{\rm TDE}}\approx0.017{\rm AU}$,
the overall average for all interactions is $\left\langle a_{{\rm pre}}\right\rangle \approx0.445{\rm AU}.$ 

\begin{table*}
\begin{tabular*}{1\columnwidth}{@{\extracolsep{\fill}}ccccc}
\hline 
outcome & Systems & Rates $\left({\rm yr^{-1}}\right)$ & Bound systems  & Unbound systems\tabularnewline
\hline 
\hline 
Collisions & $2108\ \left(0.35\%\right)$ & $1\times10^{-4}$ & $1727\ \left(82\%\right)$ & $381\ \left(18\%\right)$\tabularnewline
\hline 
$\mu$TDE & $271\ \left(0.03\%\right)$ & $2\times10^{-5}$ & $205\ \left(75.6\%\right)$ & $66\ \left(24.4\%\right)$\tabularnewline
\hline 
Tidal capture & $181\ \left(0.03\%\right)$ & $1\times10^{-5}$ & $173\ \left(95.5\%\right)$ & $5\ \left(4.5\%\right)$\tabularnewline
\hline 
\end{tabular*}

\protect\caption{\label{tab:Table_Sigma190}Same as Table \ref{tab:Summary-of-the-Results}
but with $\sigma_{{\rm kick}}=190{\rm kms^{-1}}.$ Total number of
interacting systems after NS formation is $2,560.$ The rates are
calculated per a Milky-Way galaxy.}
\end{table*}

\subsection{Gravitational wave sources and short-GRBs}

In the case of a Black Hole (BH) companion an extremely close periapsis
can result in the formation of a binary which could merge through
GW emission in a Hubble-time and produce a GW source detectable by
aLIGO. The merger itself might give rise to the accretion of the disrupted
NS on the BH and possible production of a short-GRB. 

The equation that governs the dynamics in a GW emitting systems is
given by \citet{Peters1964} 
\begin{equation}
\frac{da}{dt}=-\frac{64}{5}\frac{G^{3}m_{1}m_{2}\left(m_{1}+m_{2}\right)}{c^{5}a^{3}\left(1-e^{2}\right)^{7/2}}\left(1+\frac{73}{24}e^{2}+\frac{37}{96}e^{4}\right)
\end{equation}
\begin{equation}
\frac{de}{dt}=-e\frac{304}{15}\frac{G^{3}m_{1}m_{2}\left(m_{1}+m_{2}\right)}{c^{5}a^{4}\left(1-e^{2}\right)^{5/2}}\left(1+\frac{121}{304}e^{2}\right)
\end{equation}
where $c$ is the speed of light. Given the mass of the binary components
(for a BH companion) , the sma $a_{{\rm pSN}}$ and the binary eccentricity
$e_{{\rm pSN}}$, we can compute the merger time for each binary.
Our simulations produced $46,155$ BH-NS binaries, out of which $484$
bound systems and the rest are unbound. In Figure \ref{fig:Probabiliy_for_Interaction}
we present the number of systems that merge as a function of time
since the SN (merger time). 

\begin{figure}
\includegraphics[width=8cm]{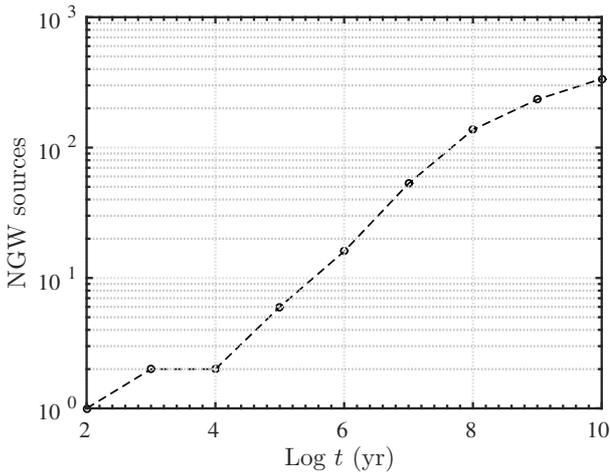}

\protect\caption{\label{fig:GWS}Cumulative number of BH-NS mergers, GW sources, as
a function of elapsed time from the SN. The post-SN merger with the
shortest delay time found in our sample occurs after only $51.6\mbox{{\rm yr}}$
after the SN explosion. }
\end{figure}

\section{Analytic understanding}
\label{sec:Analytical-understanding}
In this subsection we present our analytic treatment of the natal-kick
scenario. The binary system is comprised by a NS progenitor with mass
$m_{{\rm p}}$ that in turn becomes a NS with mass $m_{{\rm NS}}$,
the prompt mass loss is denoted by $\Delta m$, and a companion with
mass $m_{{\rm s}}$ and radius of $R_{{\rm s}}$, with a separation
$r$, sma $a_{0}$ and eccentricity $e$; for $e=0$ we get $r=a_{0}$.
We denote the combined binary mass prior the SN by $m_{{\rm b}}\equiv m_{{\rm p}}+m_{{\rm s}}$.
The combined binary mass post-SN is denoted by $m_{{\rm f}}\equiv m_{{\rm s}}+m_{{\rm NS}}$.
The companion can be of any stellar type, and therefore its size is
important for the interaction cross section with the kicked NS. 

The SN explosion governs the binary dynamical outcome through the
prompt mass loss, $\Delta m/m_{{\rm b}}$ and through the natal kick
velocity vector $\mathbf{{v_{{\rm kick}}}}$. For a detailed
treatment of sudden mass loss and natal kicks see \citet{Hills1983}.
The natal kick, $\mathbf{{v_{{\rm kick}}}}$ changes both
the energy and the angular momentum of the binary, hence it correspondingly
changes both the sma and the eccentricity. \citet{Hills1983} finds
the changes in the sma due to the natal kick and the mass loss (for
initially circular orbit) to be 
\begin{equation}
\frac{a_{{\rm pSN}}}{a}=\frac{1-\Delta m/m_{{\rm b}}}{2-2\Delta m/m_{{\rm b}}-v_{{\rm pSN}}^{2}/v_{0}^{2}},\label{eq:change_in_a}
\end{equation}
where $v_{{\rm pSN}}$ is the relative velocity of the binary components
after the SN, namely $\mathbf{\mathbf{v}_{{\rm pSN}}=\mathbf{v}_{0}+\mathbf{v}_{{\rm kick}}}$.
From the cosine theorem we get the following geometric relation 
\begin{equation}
\frac{v_{{\rm pSN}}^{2}}{v_{0}^{2}}=\left(\frac{v_{{\rm kick}}}{v_{0}}\right)^{2}+2\frac{v_{{\rm kick}}}{v_{0}}\cos\theta+1,\label{eq:V_pSN}
\end{equation}
where $\theta$ is the angle of the natal kick velocity vector and
the orbital velocity vector. From equation (\ref{eq:change_in_a})
and (\ref{eq:V_pSN}) we find a minimum value of $v_{{\rm pSN}}$
that dissociates the binary, namely for any value of $v_{{\rm pSN}}$
that satisfies the following condition
\begin{equation}
v_{{\rm pSN}}\geq v_{0}\left(1+\sqrt{2m_{{\rm f}}/m_{b}}\right),
\end{equation}
the binary is disrupted. This explains the result presented in Figure
\ref{fig:Fraction-of-bound}, that states that all binaries with initial
sma larger than $\sim1\,{\rm AU}$ are disrupted by the natal kick
received from the SN explosion.

For simplicity we consider all binaries prior the SN to be circular,
namely $e=0.$ Hence there are two relevant velocity scales in the
system that corresponds to three regimes. First, the relative orbital
velocity at the pre-SN, $v_{0}=\sqrt{Gm_{{\rm b}}/a_{0}}$. Second,
the natal kick velocity, $v_{{\rm kick}},$ drawn randomly from a
kick velocity distribution $f\left(v_{{\rm kick}}\right).$ These
velocities together with the mass loss $\Delta m$ govern the dynamics
of the system. We identify three regimes (\mbox{i}) $\tilde{v}\equiv v_{{\rm kick}}/v_{0}\gg1$
where the kick velocity is much bigger than the orbital velocity,
this implies large separation for a given kick velocity distribution,
(\mbox{ii}) $\tilde{v}\ll1$, where the kick velocity is much smaller
than the orbital velocity, this implies extremely close binary at
the moment of SN, (\mbox{iii}) $\tilde{v}\approx1$, where the kick
velocity comparable to the orbital velocity.

In the first regime $v_{{\rm kick}}\gg v_{0}$ all systems are disrupted
following the SN explosion. Further, the new relative velocity is
$v_{{\rm pSN}}\approx v_{{\rm kick}}$, and therefore in approximately
half the cases the NSs will be kicked in the direction of the companion,
namely the closest approach will be closer than the separation of
the binary in the instant of the SN. For these systems we can estimate
the cross-section for different close-interactions with the companion.
For peri-apsis passage withing a distant $R_{*}$ of the companion
we can estimate the cross section, $\sigma_{{\rm cs}}$ to be 
\begin{equation}
\sigma_{{\rm cs}}=\sigma_{{\rm geometric}}\left(1+\left(\frac{v_{{\rm esc}}}{v_{{\rm pSN}}}\right)^{2}\right),
\end{equation}
where $\sigma_{{\rm geometric}}={\rm \pi}R_{{\rm *}}^{2}$ is the
geometric cross section, and $R_{{\rm *}}$ is the relevant radius,
and the escape velocity is calculated at $R_{*}$, namely $v_{{\rm esc}}=\sqrt{G\left(m_{{\rm NS}}+m_{{\rm {\rm s}}}\right)/R_{*}}$.
We are interested in very close encounters: collisions, $\mu$TDEs
or tidal-captures. These correspond to $v_{{\rm esc}}/v_{{\rm psN}}\geq1$.
With such parameters the systems are well inside the gravitational
focusing regime, in which case the cross-section for close-interactions
scales linearly with $R_{*}$
\begin{equation}
\sigma_{{\rm cs}}\propto\sigma_{{\rm geometric}}\left(\frac{v_{{\rm esc}}}{v_{{\rm pSN}}}\right)^{2}={\rm \pi}R_{{\rm *}}^{2}\frac{G\left(m_{{\rm NS}}+m_{{\rm {\rm s}}}\right)}{R_{*}v_{{\rm pSN}}^{2}}\propto R_{*}.
\end{equation}

The fraction of the systems in which the NS encounters the companion
at a distance closer than $R_{*}$ is 
\begin{equation}
F_{1}=\int\frac{\sigma_{{\rm cs}}}{4{\rm \pi}r^{2}}{\scriptstyle f}\left(r\right)dr,\label{eq:F1}
\end{equation}
where ${\scriptstyle f}\left(r\right)$ is the separation distribution
within the relevant integration boundaries, i.e. including the separations
of all binaries. It is clear from the structure of eq. \ref{eq:F1}
that $F_{1}\propto R_{*}/r$ , i.e. the probability for a kick in
a binary with initial sma of $10{\rm AU}$ with a MS companion (taking
$R_{{\rm s}}=1R_{\odot}$) to result in a direct collision scales
like $\sim1R_{\odot}/10{\rm AU\approx4\times10^{-4}.}$

In the second regime, $v_{{\rm kick}}\ll v$, and third regime, $v_{{\rm kick}}\approx v_{0}$
most of the binaries survive the SN, this implies close pre-SN binaries
are involved. The most compact pre-SN binaries are on circular orbits
due to tidal interaction that lead to their small sma, post-common-envelope.
In this case we can simply estimate the periapsis, $q$ of the new
Keplerian orbit due to the kick and mass loss. The periapsis of the
new orbit with the sma taken from (\ref{eq:change_in_a}) 
\begin{equation}
q=a_{{\rm pSN}}\left(1-e_{{\rm pSN}}\right).
\end{equation}
The angular momentum and the energy of the binary after the SN determines
the value of $q$. We can calculate $a_{{\rm pSN}}$ from eq. (\ref{eq:change_in_a})
and can find $e_{{\rm pSN}}$ from the specific angular momentum equation,
\begin{equation}
l=\left|\mathbf{r\times}\mathbf{v_{{\rm pSN}}}\right|=\sqrt{Gm_{{\rm f}}a_{{\rm pSN}}\left(1-e_{{\rm pSN}}^{2}\right).}\label{eq:angular_momentum}
\end{equation}

We use the angles defined by \citet{Troja2010}, where $\theta$ is
the angle between the \textbf{$\mathbf{v_{0}}$ }and $\mathbf{v_{{\rm kick}}}$
and $\phi$ is the angle between the initial orbital plane and the
plane span by \textbf{$\left\{ \mathbf{v_{0}},\mathbf{v_{{\rm kick}}}\right\} $}.
Following the calculation done by \citet{Troja2010} and from eq.
(\ref{eq:angular_momentum}) we get the following conditions for a
close encounter within $q\leq R_{*}$
\begin{equation}
\sin^{2}\phi\leq\frac{\xi^{2}-\left(1+\tilde{v}\cos\theta\right)^{2}}{\tilde{v}^{2}\sin^{2}\theta}\label{eq:phi}
\end{equation}
\begin{equation}
\frac{-\left(\xi+1\right)}{\tilde{v}}\leq\cos\theta\leq\frac{\xi-1}{\tilde{v}}\label{eq:theta}
\end{equation}
\begin{equation}
\xi^{2}\equiv\frac{m_{{\rm f}}a_{{\rm pSN}}}{m_{b}a_{0}}\left[\frac{2R_{*}}{a_{{\rm pSN}}}-\left(\frac{R_{*}}{a_{{\rm pSN}}}\right)^{2}\right].\label{eq:xi}
\end{equation}

\begin{figure}
\includegraphics[width=8cm]{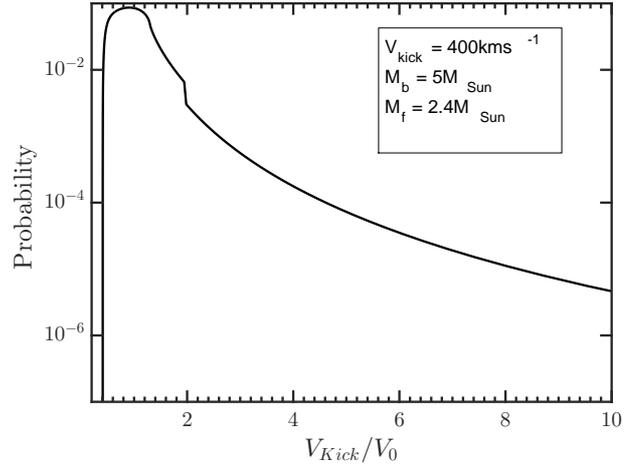}\protect\caption{\label{fig:Probabiliy_for_Interaction} The probability of interacting
of a $5\ M_{\odot}$ binary $\left(1M_{\odot}+4M_{\odot}\right)$that
undergoes a SN and becomes a $2.4\ M_{\odot}$ binary. With a natal
kick velocity of $400{\rm kms^{-1}}$ interacting at $1\ R_{\odot}.$}
\end{figure}

We note that \citet{Troja2010} approximated eq. (\ref{eq:xi}) by
$\xi=$$\left(2m_{{\rm f}}R_{*}/m_{0}a_{0}\right)^{1/2}$ while we
keep this term because we find several very close binaries prior the
SN. In Figure (\ref{fig:Probabiliy_for_Interaction}) we present the
probability for interaction between a NS and a MS star companion with
mass of $1\,{\rm M}_{\odot}$ and radius of $1\,{\rm R}_{\odot}$
at a separation of $1\ R_{\odot}$ (collision) as a function of the
$\tilde{v}$. We find that the probability distribution peaks around
unity, namely when the kick velocity approximately equals the orbital
velocity, but with the direction opposite direction in respect to
the orbital velocity; hence, the resulting angular momentum is minimized.
Consequently, we expect that in the second regime no interacting binaries
will be proceeded, while we expect most of the interacting binaries
to originate from the third regime, where $\tilde{v}\approx1.$ Indeed,
these conclusions are consistent with our numerical results shown
in Figure \ref{fig:sma-distribution-of}. We note that the difference
in comparison with the $\mu$TDE is due to the difference in the typical
companion radius, which is smaller by an order of magnitude lower,
on average, in the $\mu$TDE case when comparing with the other type
of interactions (see Figure \ref{fig:Radius_Companion}). 

\begin{figure}
\includegraphics[width=8cm]{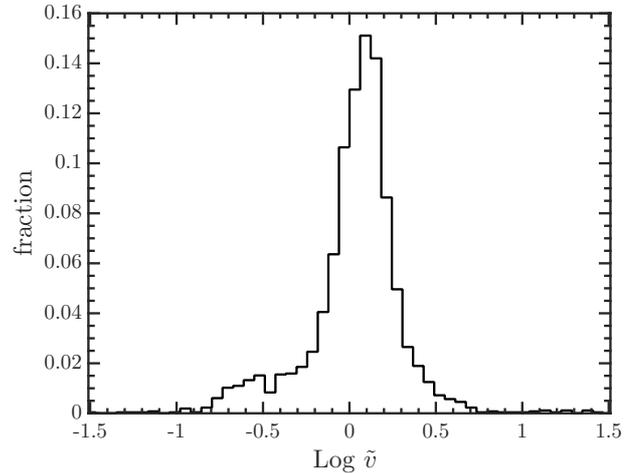}

\protect\caption{\label{fig:V_tilde}Distribution of $\tilde{v}$ for all $2,641$
interacting binaries. The distribution is peaked at the value of $\tilde{v}\approx1$,
i.e the natal kick velocity is roughly equal to the orbital velocity.}
\end{figure}

Following the analytic treatment we reach the conclusion that the
third regime is the most conductive to form a nontrivial interaction
between the components. We present the distribution of $\tilde{v}$
in Figure \ref{fig:V_tilde}; $\sim91.5\%$ of the interacting systems
have a value of $\log\tilde{v}$ between the values of $-0.5$ and
$0.5$. 

For a given binary mass in the pre-SN system and given the natal kick
velocity distribution we can define the separation scale $r_{{\rm s}}$
by the following relation: 
\begin{equation}
\left\langle v_{{\rm kick}}\right\rangle =v=\sqrt{\frac{Gm_{{\rm b}}}{r_{{\rm s}}}},\label{eq:r_c}
\end{equation}
where $\left\langle v_{{\rm kick}}\right\rangle $ is the mean kick
velocity from distribution $f\left(v_{{\rm kick}}\right)$. We note
an inverse relation between the kick velocity and the separation scale
$r_{{\rm s}}$, and therefore smaller $\left\langle v_{{\rm kick}}\right\rangle $,
corresponds to larger separation scales, which naturally explains
the results in section \ref{sub:Comparison-to-different}.

\section{Implications}
\label{sec:Implications}
Many studies explored binary stellar evolution in NS binary systems
and its outcomes and implications (see e.g. \citealt{Yoon2015}),
and in particular accounted for the effects of NS natal kicks.
These works explored the long term evolution of binaries and their
various transient outcomes (GRBs, GW sources) or the interacting X-ray
binaries or exotic objects they could produce. Nevertheless, most
of the studies did little study of the direct implications of close
encounters occurring on dynamical timescales following the SN, and
in particular did not consider binaries that formally became unbound
even if they approached sufficiently close to strongly interact immediately
after the kick. Our work focuses on these latter issues, namely the
implications of close encounters immediately after the supernova and
the natal kick, which were not, or scarcely studied before. In addition
we do consider some aspects of the longer timescale evolution, and
in particular novel implications for eLISA GWs, as well as novel electromagnetic
counterparts for GWs. 

Generally we find that in a fraction of $\sim1-3\times10^{-3}$ of
the massive systems we modeled the NSs either collided with companions;
disrupt their companions via $\mu$TDE; or formed a close binary that
in turn will eventually form an X-ray source. We note that the systems
we consider also include cases of NS-BH binaries which will merge through
GW waves inspiral in less than a Hubble time, and serve as GW sources
observable by current (aLIGO) and next generation GW detectors (eLISA).
These are extensively studied in the literature and here we only point
out novel aspects of the formation of such system and their electromagnetic counterparts. 

\textbf{$\mu$TDEs and faint peculiar SNe:} $\mu$TDEs are tidal disruption
events by stellar compact objects \citet{Perets2016} that may explain
ultra long Gamma ray bursts (GRBs). We find that $2-4\times10^{-4}$
of the NS binary systems undergo tidal disruption of the stellar companion.
The majority of the companion are found to be either He stars or WDs.
While the former may produce $\mu$TDEs as envisioned by \citet{Perets2016},
the latter disruptions of WDs by NSs might produce faint transients,
possibly resembling some types of peculiar faint SNe \citep{Metzger2012,Fernandez2013}
such as Calcium-rich SNe \citep{Perets2010a}, though further studies
of such events may shed more light on their actual observational properties.
Note that the relatively short delay times between the SN and the
$\mu$TDE would give rise to an ultra-long GRB accompanied by a SN
(likely a type Ib/c given the stripped companion), or in the case
of disruption of a WD, the initial SN might be accompanied by the
faint peculiar transient, in which case the former, regular SN might
mask the appearance of the latter.

\textbf{Short GRBs:} Our models only consider the case of a single
NS in a binary, and generally did not follow the binary later evolution
after the formation of the first NS (besides the case of a BH companion
to the newly formed NS, in which case only dynamical evolution, but
not stellar evolution is considered, as we discuss below). In principle,
if the binary survives the kick, the companion to the NS star could
itself become a NS later on, for a sufficiently massive progenitor.
In this case NS-NS interactions should be considered. \citet{Troja2010}
calculated the possibility for NS-NS collisions following natal kicks,
which may result in GW sources accompanied by short GRBs. They studied
the likelihood of such events as a function of the initial orbital
parameters, and the kick velocity (but did not consider any previous
stellar evolution leading to the pre-kick configurations). The authors
found that in order to get a direct NS-NS collision from an initially
circular orbit the kick velocity $v_{{\rm kick}}\geq v_{{\rm c}}$,
where $v_{{\rm c}}$ is the relative velocity between the NS progenitor
and its NS companion in the circular per-explosion orbit; consistent
with our analytic results.

We do consider other possible short-GRB progenitors such as BH-NS
mergers. Most interestingly, the earliest merger (due to GW inspiral)
we identify occurs only $50\,{\rm yr}$ post-SN. This suggests that
short merger times are possible. Such short merger-time events can
give rise to detectable SNe \emph{preceding} short-GRBs, where only
long-GRBs were suggested (and observed) to be connected to SNe until
now. To date no such SN-accompanied short-GRBs had been observed \citep{Nakar2007},
beside the possible marginal case of GRB 050416A (2.4 s GRB; \citep{Soderberg2007}).

\textbf{GW sources: }The possibility of short merger times for the
BH-NS case gives rise to potentially important implications for GW
sources and their electromagnetic counterparts. Sufficiently short
merger-time can result in a unique novel type of eLISA GW sources;
since eLISA is sensitive to GW sources from binaries at large separations,
it can generally track a GW binary as it inspirals from larger separations
to smaller ones for months or years (but before they enter the aLIGO
band;\citep{Sesana2016}). In the case of NS-kick - formed GW source,
the source could promptly appear at small separations, without showing
the prior typically expected longer term inspiral from larger separations,
thus providing a smoking gun signature for its natal-kick formation
process. Moreover, cases of short (<yr) merger times, if such exist,
can provide novel type of \emph{electromagnetic counterparts to aLIGO
GW sources}, i.e. a SN which \emph{precedes} (or accompanies) the
GW detection, by the delay-time time scale. All of these short merger-time
GW sources are likely to be highly eccentric when entering the detectors
observable waveband. 

Note that short merger times have been discussed in the literature
(e.g. \citealt{Belczynski2006a}), but their potential role as novel
type of GW sources and preceding electromagnetic counterparts is suggested
here for the first time, to the best of our knowledge. Note that similar
type of mergers and counterparts could be potentially be produced
in kick-induced BH-BH and NS-NS mergers which are not discussed here.
A more extensive exploration, both analytic and numeric, of these
potential novel type of GW sources and preceding electromagnetic counterpart
is beyond the scope of this paper and will be explored in another
study dedicated to these issues. 

\textbf{Physical collisions, Thorne-Zytkow objects (TZOs) and long-GRBs:}
Direct collision of a NS with a companion is likely to produce an
optical transient. The collision, and hence the transient from an
unbound systems will occur with a distance of $r$, the separation
between the components at the moment of SN, with $r\approx1.07{\rm AU},$
with for the bound system $r\approx0.3{\rm AU}.$ This transient is
triggered after an average time of $\left\langle \Delta t\right\rangle _{{\rm collision}}\approx4.7\times10^{6}{\rm sec}$
with $\left\langle \Delta t\right\rangle _{{\rm collision,bound}}\approx6.17\times10^{6}{\rm sec}$
for the bound systems and $\left\langle \Delta t\right\rangle _{{\rm collision,unbound}}\approx1.17\times10^{6}{\rm sec}$
for the unbound systems. All values of the elapsed time corresponds
to $\sim10-80\,{\rm days}$. Any transient due to such collisions
is therefore like to occur while the SN (that produced the NS and
its kick) is still observable. In most cases such SNe would likely
be type Ib/c as the NS progenitor is usually a stripped star. Such
collisions could also have long-term implications in producing Thorn-Zytkow
objects (TZOs; \citealt{Thorne1977}), stars with degenerate neutron
core, as discussed by \citep{Leonard1994}. \citet{Brandt1995} used
a Monte Carlo simulations to investigate the effects of high SN kick
velocities on the binaries post NS formation. The authors focused
on high mass and low mass X-ray sources and the probability for a
merger in order to become a TZO. They found TZO formation rate through
this channel (see \citet{Podsiadlowski1995} for other possible channels)
to be $\sim3\times10^{-3}{\rm yr^{-1}}$ . In the context of our work
we found that $\sim9\times10^{-5}{\rm yr^{-1}}$ of the systems results
in collisions of NSs and massive ($>8M_{\odot}$) main-sequence companions,
which can considered as potential candidates for TZOs, if such objects
can exist. If instead the NS accretes the star with which it collides,
it might collapse into a BH, possibly producing a long-GRB. In this
case one might expect a long-GRB which is preceded by the SN, rather
that the typical case discussed in the literature where the initial
collapse produced the long GRB. 

\textbf{X-ray binaries:} X-ray binaries with a NS primary are binary
systems with short period orbits allowing for mass transfer from the
primary, usually evolved star, to an accretion disk around the NS.
We find that $\sim1-3\times10^{-4}$ of the NSs end up in sufficiently
close eccentric orbits as to become X-ray binaries either immediately
after the SN or later on when the companion evolves into its giant
phase and Roche-lobe overflow commences. Out of $156$ NSs that are
captured into a close orbit ($181$systems for $\sigma_{{\rm kick}}=190{\rm kms^{-1})}$
$11$ host low MS ($<2M_{\odot}$) companions; progenitors for a low
mass X-ray binaries systems (LMXBs), while the rest are more massive
progenitors producing high Mass X-ray binaries (HMXBs), usually on
a wider orbit than the LMXBx.

\section{Summary}
\label{sec:Discussion-and-Summary}
In this work we have explored the end result of natal-kicks given
to NS in binary systems. We identified three regimes corresponding
to the value of the kick velocity, and the importance of the mass
loss, $\Delta m$ and sma $a$. We preformed population synthesis
of $587,019$ binary systems and evolved them until NS is formed.
Next, we randomized the position in the binary orbit and drawn a natal
kick from a Maxwellian velocity distribution with $\sigma_{{\rm kick}}=270{\rm kms^{-1}}$
($\sigma_{{\rm kick}}=190{\rm kms^{-1}}$). Using eq. (\ref{eq:minimal r})
we determined the closet approach $r_{{\rm min}}$ of any NS to its
companion, whether the binary survived the SN or whether it was disrupted.
From the value of $r_{{\rm min}}$ and the companion radius, $R_{{\rm s}}$
and mass we identified the systems that underwent a strong encounter:
(\mbox{i}) direct collision, namely $r_{{\rm min}}<R_{{\rm s}}+R_{{\rm NS}}$
(\mbox{ii}) tidal disruption event, namely $R_{{\rm s}}+R_{{\rm NS}}<r_{{\rm min}}<R_{{\rm s}}\left(2m_{{\rm NS}}/m_{{\rm s}}\right)^{1/3}$
(\mbox{iii}) tidal capture, satisfying the condition in eq. (\ref{eq:Tidal_Capture_Criteria})
and not flagged as collision or $\mu$TDE. 

Using a simple analytic treatment we identified the important parameters
that determine the natal-kick outcomes, and find them to explain well
the results qualitatively . 

Our calculations provide us for estimate of the rate of various type
of transient events and possible production of exotic stars induced
by the natal kicks, which are discussed in details. In particular
we calculate the rates of $\mu$TDEs as well as faint SNe from accretion
of WDs on NSs; and we find the production rates of Thorn-Zytkow objects
(if such exist; or the possible production of long-GRBs if the NS
rapidly accrete the star with which they collide to collapse into
a BH) and NS X-ray binaries (the latter were also studied by others;
e.g. \citealt{Brandt1995,Podsiadlowski1995,Belczynski2006a}). Finally
we consider the possible production of short GRBs as well GW sources
from BH-NS mergers and point out the possible existence of short merger
times since the SN; such cases could give rise to novel types of eLISA
GW sources, as well as provide SN electromagnetic counterparts \emph{preceding}
aLIGO GW sources from BH-NS mergers.

\section*{Acknowledgements}
We thank the Israel science foundation
excellence center I-CORE grant 1829/12.

\bibliographystyle{mnras}
\bibliography{triple}

\end{document}